\documentclass[aps,prd,onecolumn,groupedaddress,showpacs,nofootinbib,amssymb, preprint]{revtex4-2}
\usepackage[dvips]{graphicx}
\usepackage{amssymb}
\usepackage{amsmath}
\usepackage{graphicx,,color}
\usepackage{amsfonts}
\usepackage{bm}
\usepackage{cancel}
\usepackage{comment}
\usepackage{floatflt}
\usepackage{slashed}
\allowdisplaybreaks[4]
\begin{document}

\preprint{KEK-TH-2690, KEK-Cosmo-0373}
\title{
The correspondence of generalised entropic cosmology theory with $F(T)$ and $F(Q)$ modified gravity and gravitational waves
}

\author{Shin'ichi~Nojiri$^{1,2}$}\email{nojiri@gravity.phys.nagoya-u.ac.jp}
\author{Sergei~D.~Odintsov$^{3,4}$}\email{odintsov@ice.csic.es}

\affiliation{ $^{1)}$ Theory Center, High Energy Accelerator Research Organization (KEK), \\
Oho 1-1, Tsukuba, Ibaraki 305-0801, Japan \\
$^{2)}$ Kobayashi-Maskawa Institute for the Origin of Particles and the Universe, \\
Nagoya University, Nagoya 464-8602, Japan \\
$^{3)}$ Institute of Space Sciences (ICE, CSIC) \\ 
C. Can Magrans s/n, 08193 Barcelona, Spain \\
$^{4)}$ ICREA, Passeig Lluis Companys, 23, 08010 Barcelona, Spain
}

\begin{abstract}

We investigate the correspondence between modified gravity theories and general entropic cosmology theory. 
Such a theory is proposed by an analogy with Jacobson's work, where the Einstein equation was derived from the Bekenstein-Hawking entropy. 
We compare FLRW equations obtained in entropic gravity with those in modified gravity theories. 
It is found the correspondence of $F(T)$ and $F(Q)$ gravities and general entropic gravity. 
We regard the $F(T)$ and $F(Q)$ gravity theories as effective local theories corresponding to the entropic gravity theories 
and we investigate the gravitational waves. 
The obtained equation of the gravitational wave is identical to that in Einstein's gravity except that the gravitational coupling is modified 
by the functional form of the functions $F(T)$ and $F(Q)$. 
It is interesting that in the case of the Tsallis entropic cosmology, the gravitational coupling becomes small or large, which may suppress or enhance 
the emission of the gravitational wave. 

\end{abstract}

\maketitle

\newpage

\section{Introduction}

As well-known, the thermodynamics of black holes started with Bekenstein-Hawking entropy~\cite{Bekenstein:1973ur, Hawking:1975vcx},
\begin{align}
S = \frac{A}{4G} \, .
\label{BH-entropy}
\end{align}
Here $A = 4\pi r_\mathcal{h}^2$ is the area of the horizon given by the horizon radius $r_\mathcal{h}$. 
The Bekenstein-Hawking entropy was widely applied in the study of black hole thermodynamics and entropic cosmology related to the apparent horizon.

On the other side, in different physical and statistical systems, there exist various entropies.
These are following ones: Tsallis entropy~\cite{Tsallis:1987eu}, R\'{e}nyi entropy~\cite{Renyi}, 
Barrow entropy~\cite{Barrow:2020tzx}, Sharma-Mittal entropy~\cite{SayahianJahromi:2018irq}, 
Kaniadakis entropy~\cite{Kaniadakis:2005zk, Drepanou:2021jiv}, and Loop Quantum Gravity entropy~\cite{Majhi:2017zao, Liu:2021dvj}. 
There appeared a number of papers where modified FRLW equations based on the above entropies were derived. 
The corresponding early- and late-time cosmology was investigated. 
In this way, the entropic cosmology (or entropic gravity) has been investigated.

Recently, the authors proposed four- and six-parameter generalised entropies~\cite{Nojiri:2022aof, Nojiri:2022dkr}. 
All the known above entropies %in (\ref{BH-entropy}), (\ref{Tsallis entropy}), (\ref{Renyi entropy}), (\ref{Barrow-entropy}), (\ref{SM entropy}), (\ref{K-entropy}), and (\ref{LQG entropy}) 
are just particular examples of generalised entropy in some limits of the parameters. 
In addition, three-parameter generalised entropy~\cite{Nojiri:2022aof} and five-parameter entropy were proposed in \cite{Odintsov:2022qnn}, 

Some time ago, Jacobson has shown that the Einstein equation can be obtained from the Bekenstein-Hawking entropy by the thermodynamical 
consideration~\cite{Jacobson:1995ab}. 
In the same way, one can get modified FLRW equations for different entropies (entropic cosmology).
However, such an approach lacks the Lagrangian description which causes a number of problems. 
For instance, it is obscure how to describe gravitational waves (GWs) in entropic cosmology. 
One natural possibility is to search for the correspondence between entropic gravity and modified gravity. 
In this way, one can get effective lagrangian description for entropic cosmology.
Then a natural question is what kind of gravity theories correspond to the above general entropic cosmology? 
The candidates should be searched among modified gravities, see reviews~\cite{Nojiri:2010wj, Capozziello:2011et, Nojiri:2017ncd}. 
This paper is devoted to the resolution of this problem. 

The keys to solving the above problem could be obtained from the cosmological equations. 
By applying the Bekenstein-Hawking entropy for a cosmological horizon and by using the thermodynamical considerations, it has been shown 
that the first Friedmann-Lema\^{i}tre-Robertson-Walker (FLRW) equation, which is derived from the Einstein equation and describes the cosmology, 
can be obtained~\cite{Padmanabhan:2009vy, Cai:2005ra}. 
If other entropy formulas are used, the first FLRW equation is modified in a way that reminds us of the alternative gravity. 
Then if we find any gravity theory reproducing the same modified first FLRW equation, the theory gives the correspondence with that based on specific entropy. 
Furthermore, because modified gravity is based on the local Lagrangian density, we can discuss the dynamical phenomena like gravitational wave 
corresponding to the generalised entropic cosmology. 

This paper is constructed as follows. 
In the next section, the review of the entropy functions proposed so far is done. 
We explain how the first FLRW equation is modified by the general entropies in Section~\ref{SecIII}. 
We then show that the gravity theory reproducing the modified entropic first FLRW equation is $F(T)$ gravity (Section~\ref{FTcor}) 
or $F(Q)$ gravity (Section~\ref{FQrorr}). 
It is also shown that it is hard to find the correspondence by using $F(R)$ or Einstein-$F(\mathcal{G})$ gravity. 
Here $\mathcal{G}$ is the Gauss-Bonnet invariant. 
In Section~\ref{SecGW}, we review the gravitational waves study in the framework of $F(T)$ and $F(Q)$ gravities. 
In both gravity theories, at least perturbatively, the only propagating mode is a massless and spin-two gravitational wave. 
This situation does not change from the gravitational wave in Einstein's gravity. 
The gravitational coupling is, however, changed from that in Einstein's gravity. 
Then we speculate on the possibility that the gravitational wave is enhanced or weakened in the early universe. 
The last section is devoted to summary and discussion. 

\section{A generalised entropies}\label{sec-gen-entropy}

Many kinds of entropies have been proposed after Bekenstein-Hawking entropy (\ref{BH-entropy})~\cite{Bekenstein:1973ur, Hawking:1975vcx}. 
Tsallis entropy~\cite{Tsallis:1987eu} looks as 
\begin{align}
S_\mathrm{T} = \frac{A_0}{4G}\left(\frac{A}{A_0}\right)^{\delta} \, .
\label{Tsallis entropy}
\end{align}
Here $A_0$ is a constant with the dimension of area and $\delta$ is the exponent. 
The R\'{e}nyi entropy is given by \cite{Renyi},
\begin{align}
S_\mathrm{R} = \frac{1}{\alpha} \ln \left( 1 + \alpha S \right) \, .
\label{Renyi entropy}
\end{align}
This expression (\ref{Renyi entropy}) includes a parameter $\alpha$ and $S$ is Bekenstein-Hawking entropy in (\ref{BH-entropy}). 
Barrow proposed the following entropy~\cite{Barrow:2020tzx},
\begin{align}
\label{Barrow-entropy}
S_\mathrm{B} = \left(\frac{A}{A_\mathrm{Pl}} \right)^{1+\Delta/2} \, .
\end{align}
Here $A$ is the area of the black hole horizon as in (\ref{BH-entropy}) and $A_\mathrm{Pl} = 4G$ is called the Planck area. 
The Sharma-Mittal entropy is given by \cite{SayahianJahromi:2018irq},
\begin{align}
S_\mathrm{SM} = \frac{1}{R}\left[\left(1 + \delta ~S\right)^{R/\delta} - 1\right] \, .
\label{SM entropy}
\end{align}
The expression (\ref{SM entropy}) includes two parameters $R$ and $\delta$. 
The Kaniadakis entropy function with a phenomenological parameter $K$ 
has the following form \cite{Kaniadakis:2005zk, Drepanou:2021jiv}, 
\begin{align}
S_\mathrm{K} = \frac{1}{K}\sinh{\left(KS\right)} \, .
\label{K-entropy}
\end{align}
In the context of Loop Quantum Gravity, the following entropy function has also been proposed \cite{Majhi:2017zao, Liu:2021dvj}
\begin{align}
S_q = \frac{1}{\left(1-q\right)}\left[\mathrm{e}^{(1-q)\Lambda(\gamma_0)S} - 1\right] \, .
\label{LQG entropy}
\end{align}
Here $q$ is the exponent and $\Lambda(\gamma_0) = \ln{2}/\left(\sqrt{3}\pi\gamma_0\right)$ with $\gamma_0$ is called the Barbero-Immirzi parameter. 

After that, the following four- and six-parameter generalised entropies~\cite{Nojiri:2022aof, Nojiri:2022dkr} were proposed,
\begin{align}
S_4 \left(\alpha_{\pm},\delta,\gamma\right) = \frac{1}{\gamma}\left[\left(1 + \frac{\alpha_+}{\delta} S\right)^{\delta}
 - \left(1 + \frac{\alpha_-}{\delta} S\right)^{-\delta}\right] \, ,
\label{intro-1}
\end{align}
and
\begin{align}
S_6 \left(\alpha_{\pm},\delta_{\pm},\gamma_{\pm}\right) = \frac{1}{\alpha_+ + \alpha_-}
\left[ \left( 1 + \frac{\alpha_+}{\delta_+} S^{\gamma_+} \right)^{\delta_+} 
 - \left( 1 + \frac{\alpha_-}{\delta_-} S^{\gamma_-} \right)^{-\delta_-} \right]\, ,
\label{intro-2}
\end{align}
respectively. 
For some parameter choices, as particular examples, the above two entropies describe all the known entropies, (\ref{BH-entropy}), (\ref{Tsallis entropy}), 
(\ref{Renyi entropy}), (\ref{Barrow-entropy}), (\ref{SM entropy}), (\ref{K-entropy}), and (\ref{LQG entropy}). 
In addition, an entropy with three parameters was also proposed in \cite{Nojiri:2022aof}, 
\begin{align}
S_3\left(\alpha,\delta,\gamma\right) = \frac{1}{\gamma}\left[\left(1 + \frac{\alpha}{\delta} S\right)^\delta - 1\right] \, .
\label{intro-3}
\end{align}
$S_3$ gives all the above entropies except the Kaniadakis entropy. 
Furthermore, to solve the problem of singularity, which occurs at the bouncing time when $H=0$ if we consider the bouncing universe, 
the following entropy with five parameters was proposed in \cite{Odintsov:2022qnn}, 
\begin{align}
S_5\left(\alpha_{\pm},\delta,\gamma,\epsilon\right) 
= \frac{1}{\gamma}\left[\left\{1 + \frac{1}{\epsilon}\tanh\left(\frac{\epsilon \alpha_+}{\delta} S\right)\right\}^{\delta}
 - \left\{1 + \frac{1}{\epsilon}\tanh\left(\frac{\epsilon \alpha_-}{\delta} S\right)\right\}^{-\delta} \right] \, .
\label{intro-4}
\end{align}
The $\tanh$ function makes the entropy (\ref{intro-4}) to be non-singular due to the universe's evolution starting from the Big Bang. 

One may postulate that the minimum number of parameters required in a generalised entropy function which generalises 
all the known entropies in (\ref{BH-entropy}), (\ref{Tsallis entropy}), 
(\ref{Renyi entropy}), (\ref{Barrow-entropy}), (\ref{SM entropy}), (\ref{K-entropy}), and (\ref{LQG entropy}) is equal to four.

\section{Generalised entropies and cosmology}\label{SecIII}

Let us now consider the cosmology described by generalised entropies. 

FLRW spacetime with a flat spacial part is given by metric, 
\begin{align}
ds^2=-dt^2+a^2(t)\sum_{i=1,2,3} \left(dx^i\right)^2 \, .
\label{metric}
\end{align}
Here $a(t)$ is a scale factor. 

The radius $r_\mathrm{H}$ of the cosmological horizon is given by, 
\begin{align}
\label{apphor}
r_\mathrm{H}=\frac{1}{H}\, .
\end{align}
Hereth $H = \dot{a}/a$ is the Hubble rate. 
The entropy contained inside the cosmological horizon is obtained by using 
the Bekenstein-Hawking relation \cite{Padmanabhan:2009vy}. 
The flux of the energy $E$, which is the increase of the heat $Q$ in the region, is given by, 
\begin{align}
\label{Tslls2}
dQ = - dE = -\frac{4\pi}{3} r_\mathrm{H}^3 \dot\rho dt = -\frac{4\pi}{3H^3} \dot\rho dt 
= \frac{4\pi}{H^2} \left( \rho + p \right) dt \, .
\end{align}
We used the conservation law: $0 = \dot \rho + 3 H \left( \rho + p \right)$ in the last equality. 
As the Hawking temperature is given by \cite{Cai:2005ra}, 
\begin{align}
\label{Tslls6}
T = \frac{1}{2\pi r_\mathrm{H}} = \frac{H}{2\pi}\, ,
\end{align}
by also using the first law of thermodynamics $TdS = dQ$, 
we obtain $\dot H = - 4\pi G \left( \rho + p \right)$. 
Integration of the expression leads to the first FLRW equation, 
\begin{align}
\label{Tslls8}
H^2 = \frac{8\pi G}{3} \rho + \frac{\Lambda}{3} \, .
\end{align}
Here the constant of the integration $\Lambda$ is regarded as a cosmological constant. 

Instead of the Bekenstein-Hawking entropy of Eq.~(\ref{BH-entropy}), we may use the generalised entropy $S_\mathrm{g}$. 
Then the first law of thermodynamics leads to the following equation, 
\begin{align}
\dot{H}\left. \left(\frac{\partial S_\mathrm{g}}{\partial S}\right) \right|_{S = \frac{A}{4G} = \frac{\pi}{GH^2}}= -4\pi G\left(\rho + p\right) \,.
\label{FRW1-subB}
\end{align}
By using the conservation law of the matter $\dot{\rho} + 3H\left(\rho + p\right) = 0$, again, 
Eq.~(\ref{FRW1-subB}) can be rewritten as,
\begin{align}
\label{SgH}
\left. \left(\frac{\partial S_\mathrm{g}}{\partial S}\right) \right|_{S = \frac{A}{4G} = \frac{\pi}{GH^2}} d\left( H^2 \right) = \left(\frac{8\pi G}{3}\right)d\rho \,. 
\end{align}
By integrating (\ref{SgH}), we obtain a generalised first FLRW equation, 
\begin{align}
\label{gSeq}
\mathcal{H} \left( H^2 \right)^2 \equiv 
\int^{H^2} dx \left. \left(\frac{\partial S_\mathrm{g}}{\partial S}\right) \right|_{S = \frac{\pi}{Gx}} = \left(\frac{8\pi G}{3}\right)\rho \,. 
\end{align}
Especially in the case of four-parameter generalised entropies in (\ref{intro-1}), we obtain, 
\begin{align}
\frac{GH^4\beta}{\pi\gamma}&\,\left[ \frac{1}{\left(2+\beta\right)}\left(\frac{GH^2\beta}{\pi\alpha_-}\right)^{\beta}
{}_2F_{1}\left(1+\beta, 2+\beta, 3+\beta, -\frac{GH^2\beta}{\pi\alpha_-}\right) \right. \nonumber\\ 
&\, \left. + \frac{1}{\left(2-\beta\right)}\left(\frac{GH^2\beta}{\pi\alpha_+}\right)^{-\beta}
{}_2F_{1}\left(1-\beta, 2-\beta, 3-\beta, -\frac{GH^2\beta}{\pi\alpha_+}\right) \right] 
= \frac{8\pi G\rho}{3} + \frac{\Lambda}{3} \, .
\label{FRW-2}
\end{align}
Here $\Lambda$ is the integration constant corresponding to the cosmological constant. 
$_2F_1$ is the Hypergeometric function. 
By considering proper parameter choices, we obtain the first FLRW equations corresponding to the entropies in (\ref{BH-entropy}), (\ref{Tsallis entropy}), 
(\ref{Renyi entropy}), (\ref{Barrow-entropy}), (\ref{SM entropy}), (\ref{K-entropy}), and (\ref{LQG entropy}). 

\begin{itemize}
\item In the case of the Tsallis entropy~(\ref{Tsallis entropy}) or the Barrow entropy~(\ref{Barrow-entropy}), 
by identifying $\beta = \delta$ or $\beta = 1 + \Delta$, respectively, first we choose $\alpha_- \to 0$. 
After that, by choosing $\gamma = \left(\alpha_+/\beta\right)^{\beta}$ and keeping $\beta$ finite, we consider the limit of $\alpha_+ \rightarrow \infty$. 
Then we obtain the Tsallis entropy~(\ref{Tsallis entropy}) or the Barrow entropy~(\ref{Barrow-entropy}). 
In the procedure of the limits, because 
\begin{align}
\label{hgf}
{}_2F_1 \left( \alpha, \beta, \gamma, z \right) 
=&\, \frac{\Gamma(\gamma) \Gamma(\beta - \alpha)}{\Gamma(\beta) \Gamma(\gamma - \alpha)} \left( - z \right)^{-\alpha} 
F \left( \alpha, \alpha + 1 - \gamma, \alpha + 1 - \beta, \frac{1}{z} \right) \nonumber \\
&\, + \frac{\Gamma(\gamma) \Gamma(\alpha - \beta)}{\Gamma(\alpha) \Gamma(\gamma - \beta)} \left( - z \right)^{-\beta} 
F \left( \beta, \beta + 1 - \gamma, \beta + 1 - \alpha, \frac{1}{z} \right) \, , \nonumber \\
{}_2F_1 \left( \alpha, \beta, \gamma, 0 \right) =&\, 1\, ,
\end{align}
we obtain
\begin{align}
\label{HTB}
\mathcal{H} \left( H^2 \right)^2 
= \frac{\beta}{2 - \beta} \left(\frac{GH^2}{\pi}\right)^{1 -\beta} { H^2} \, ,
\end{align}
which coincides with the result in the previous works (see \cite{Nojiri:2022aof}, for example). 
\item The R\'{e}nyi entropy (\ref{Renyi entropy}) can be obtained in the limit $\alpha_- = 0$, $\beta \rightarrow 0$, 
and $\alpha \equiv \frac{\alpha_+}{\beta} \rightarrow \mathrm{finite}$ 
and by choosing $\gamma = \alpha_+$. 
Therefore Eq.~(\ref{FRW-2}) gives, 
\begin{align}
\label{HTR}
\mathcal{H} \left( H^2 \right)^2 
= \frac{GH^4}{2 \pi \alpha} {}_2F_1 \left( 1,2,3, - \frac{GH^2}{\pi\alpha} \right)\, .
\end{align}
Then we find the cosmology which corresponds to R\'{e}nyi entropy (\ref{Renyi entropy}). 
\end{itemize}
Similarly, 
\begin{itemize}
\item The Sharma-Mittal entropy (\ref{SM entropy}) is obtained in the limit of $\alpha_- \to 0$ 
by identifying $\gamma = R$, $\alpha_+ = R$, and $\beta = R/\delta$.
\item In the limit of $\beta \rightarrow \infty$, by identifying $\alpha_+ = \alpha_- = \frac{\gamma}{2} = K$, we 
obtain the Kaniadakis~entropy (\ref{K-entropy}). 
\item The Loop Quantum Gravity entropy (\ref{LQG entropy}) with $\Lambda(\gamma_0) = 1$ or equivalently $\gamma_0 = \frac{\ln{2}}{\pi\sqrt{3}}$ is obtained 
in the limit $\alpha_- = 0$, by identifying $\beta \rightarrow \infty$ and $\gamma = \alpha_+ = (1-q)$.
\end{itemize}

{ 
In \cite{Saridakis:2020lrg}, the gravity-thermodynamics conjecture has been applied in the case of Barrow entropy 
and similar analysis has been applied in the case of Kaniadakis entropy in \cite{Lymperis:2021qty}. 
}
In the following, we consider what kind of model with the local Lagrangian density can reproduce the generalised first FLRW equation (\ref{gSeq}). 
It is difficult to reproduce (\ref{gSeq}) by modified gravity theories based on curvature like $F(R)$ gravity or Einstein-Gauss-Bonnet gravity. 
Nevertheless, we show that $F(T)$ gravity based on the torsion and $F(Q)$ gravity based on the non-metricity can do it. 

\section{Correspondence between $F(T)$ gravity and generalised entropies}\label{FTcor}

The ``teleparallelism'' with the Weitzenb\"{o}ck connection is a gravitational theory alternative to Einstein's general relativity. 
In teleparallel gravity, a scalar quantity $T$ called torsion is used as a fundamental ingredient instead of the curvature $R$, 
which is defined by the Levi-Civita connection. 
See \cite{Hehl:1976kj, Hayashi:1979qx, Cai:2015emx, Flanagan:2007dc, Maluf:2013gaa, Garecki:2010jj, Ferraro:2008ey} for detail. 
With the motivation to explain both the inflationary era and the accelerating expansion of the current Universe, 
there has been proposed a model whose Lagrangian density is extended to be a function of $T$ as $F(T)$ 
\cite{Bengochea:2008gz, Maluf:2013gaa, Linder:2010py, Cai:2011tc, Chen:2010va, Bamba:2010wb, 
Zhang:2011qp, Nashed:2014uta, Geng:2011aj, Boehmer:2012uyw, Nashed:2014lva, Gonzalez:2011dr, Karami:2012fu, Bamba:2012vg, 
Rodrigues:2012qua, Capozziello:2012zj, Chattopadhyay:2012eu, Izumi:2012qj, Li:2013xea, Ong:2013qja, Otalora:2013tba, 
Nashed:2013bfa, Kofinas:2014owa, Harko:2014sja, ElHanafy:2014efn, Junior:2015bva, Ruggiero:2015oka, ElHanafy:2014tzj, Nunes:2016plz} 
as in $F(R)$ gravity. 
However, a crucial problem in $F(T)$ gravity is that the model includes superluminal propagating modes, which appear non-perturbatively \cite{Izumi:2012qj, Ong:2013qja}. 
The existence of the superluminal mode tells us that $F(T)$ gravity cannot be regarded as a physically consistent theory. 
Therefore the $F(T)$ gravity theory could be an effective theory for more fundamental theory at best. 

In this paper, we use the orthonormal tetrad components $e_A (x^{\mu})$. 
Here $A$ is an index of the tangent spacetime at each point $x^{\mu}$ of the spacetime manifold 
and $e_A^\mu$ forms the tangent vector of the manifold $\left( A=0,\, 1,\, 2,\, 3\right)$. 
The metric $g^{\mu\nu}$ is given by $g_{\mu\nu}=\eta_{A B} e^A_\mu e^B_\nu$ by using the flat metric $\eta_{AB}$ of the tangent spacetime. 
${T^\rho}_{\mu\nu}$ and contorsion ${K^{\mu\nu}}_\rho$ tensors are defined by,
\begin{align}
\label{eq:2.2}
T^\rho_{\ \mu\nu} \equiv e^\rho_A \left( \partial_\mu e^A_\nu - \partial_\nu e^A_\mu \right)\, , \quad 
{K^{\mu\nu}}_\rho \equiv -\frac{1}{2} \left({T^{\mu\nu}}_\rho - {T^{\nu \mu}}_\rho - {T_\rho}^{\mu\nu}\right)\, .
\end{align}
Furthermore, the torsion scalar $T$ is defined by \cite{Hayashi:1979qx,Maluf:2013gaa},
\begin{align}
\label{eq:2.4}
T \equiv {S_\rho}^{\mu\nu} {T^\rho}_{\mu\nu}\, , \quad 
{S_\rho}^{\mu\nu} \equiv \frac{1}{2} \left({K^{\mu\nu}}_\rho + {\delta^\mu}_\rho {T^{\alpha\nu}}_{\alpha} - {\delta^\nu}_\rho {T^{\alpha \mu}}_\alpha \right)\, .
\end{align}
By using $T$, the action of the $F(T)$ gravity \cite{Linder:2010py} is given by, 
\begin{align}
\label{eq:2.6}
S = \int d^4x |e| \left[ \frac{F(T)}{2{\kappa}^2}
+ \mathcal{L}_\mathrm{matter} \right]\, .
\end{align}
Here $\kappa^2=8\pi G$, 
$|e|= \det \left(e^A_\mu \right)=\sqrt{-g}$, and $\mathcal{L}_\mathrm{matter}$ is the Lagrangian density of matter.
By the variation of the action~(\ref{eq:2.6}) with respect to the tetrad field $e_A^\mu$, we obtain \cite{Bengochea:2008gz}, 
\begin{align}
\label{eq:2.7}
\frac{1}{e}
\partial_\mu \left( e{S_A}^{\mu\nu} \right) F' - e_A^\lambda {T^\rho}_{\mu\lambda} {S_\rho}^{\nu\mu} F' 
+ {S_A}^{\mu\nu} \partial_\mu T F'' +\frac{1}{4} e_A^\nu F 
= \frac{\kappa^2}{2} e_A^\rho {T_{\mathrm{matter}\, \rho}}^\nu \, .
\end{align}
Here ${T_{\mathrm{matter}\, \rho}}^\nu$ is the energy-momentum tensor of the matter. 

By choosing the FLRW metric with a flat spatial part (\ref{metric}), the tetrad components 
are given by $e^A_\mu = \mathrm{diag} \left(1,a,a,a \right)$, which yields $g_{\mu \nu}= \mathrm{diag} \left(-1, a^2, a^2, a^2 \right)$, 
which give the torsion scalar $T$ as $T=-6H^2$. 
Then Eq.~(\ref{eq:2.7}) takes the following form,
\begin{align}
\label{FTfirst}
\frac{1}{6} \left. \left( f \left( T \right) - 2T f' \left( T \right) \right) \right|_{T=-6H^2}= \left(\frac{8\pi G}{3}\right) \rho \, .
\end{align}
We should note that the torsion scalar $T$ is given by $T=-6H^2$ in the flat FRW background (\ref{metric}). 
Therefore the l.h.s. of Eq.~(\ref{FTfirst}) is a function of $H^2$ and does not depend on $\dot H$, $\ddot H$, etc. 
By comparing Eq.~(\ref{FTfirst}) with (\ref{gSeq}), we find 
\begin{align}
\label{gSeqFT}
\frac{1}{6} \left. \left( f \left( T \right) - 2T f' \left( T \right) \right) \right|_{T=-6H^2}= \mathcal{H} \left( H^2 \right)^2 \equiv 
\int^{H^2} dx \left. \left(\frac{\partial S_{g}}{\partial S}\right) \right|_{S = \frac{\pi}{Gx}} \, . 
\end{align}
Because $f \left( T \right) - 2T f' \left( T \right) = 2 \left( - T \right)^\frac{3}{2} \left( \left( - T \right)^{-\frac{1}{2}} f \left( T \right) \right)'$, 
we can integrate Eq.~(\ref{gSeqFT}) and obtain 
\begin{align}
\label{fT}
F(T) =&\, 3 \left( - T \right)^\frac{1}{2} \int^{-\frac{T}{6}} dy \left( 6 y \right)^{- \frac{3}{2}} \mathcal{H} \left( y \right)^2 \nonumber \\
=&\, 3 \left( - T \right)^\frac{1}{2} \int^{-\frac{T}{6}} dy \left( 6 y \right)^{- \frac{3}{2}} 
\int^y dx \left. \left(\frac{\partial S_\mathrm{g}}{\partial S}\right) \right|_{S = \frac{\pi}{Gx}} \,. 
\end{align} 
Therefore we found $F(T)$ gravity corresponding to the cosmology based on generalised entropy $S_\mathrm{g}(S)$. 

In the case of four-parameter generalised entropies in (\ref{intro-1}), by using (\ref{FRW-2}), we find, 
\begin{align}
\label{FRW-2FT}
F(T) =&\, \frac{3G\beta}{\pi\gamma} \left( - T \right)^\frac{1}{2} \int^{-\frac{T}{6}} dy \left( 6 \right)^{- \frac{3}{2}} y^\frac{1}{2}
\left[ \frac{1}{\left(2+\beta\right)}\left(\frac{G\beta y}{\pi\alpha_-}\right)^{\beta}
{}_2F_{1}\left(1+\beta, 2+\beta, 3+\beta, -\frac{G\beta y}{\pi\alpha_-}\right) \right. \nonumber\\ 
&\, \left. + \frac{1}{\left(2-\beta\right)}\left(\frac{G\beta y}{\pi\alpha_+}\right)^{-\beta}
{}_2F_{1}\left(1-\beta, 2-\beta, 3-\beta, -\frac{G\beta y}{\pi\alpha_+}\right) \right] \, .
\end{align}
By taking several limits for the parameters $\alpha_\pm$, $\beta$, and $\gamma$, 
we obtain the expressions of $F(T)$ for different entropies. 

For an example, for the Tsallis entropy~(\ref{Tsallis entropy}) or the Barrow entropy~(\ref{Barrow-entropy}), 
by using (\ref{HTB}), we find, 
\begin{align}
\label{HTBFT}
F(T) = - \frac{36 \beta}{\left(2 - \beta\right)\left( 3 + 2\beta\right)} \left(\frac{G}{\pi}\right)^{1-\beta} \left( - T \right)^{1-\beta} 
+ C \left( - T \right)^\frac{1}{2} \, .
\end{align}
Here $C$ is a constant of the integration. 
We should note that there appear the fractional powers of $T$. 

We also obtain the expression of $F(T)$ for the R\'{e}nyi entropic cosmology (\ref{Renyi entropy}) by using (\ref{HTR}) as follows, 
\begin{align}
\label{HTRFT}
F(T) =&\, 3 \left( - T \right)^\frac{1}{2} \int^{-\frac{T}{6}} dy \left( 6 y \right)^{- \frac{3}{2}} 
\frac{Gy^2}{2 \pi \alpha} {}_2F_1 \left( 1,2,3, - \frac{Gy}{\pi\alpha} \right)\, .
\end{align}
Similarly, we can obtain the expressions of $F(T)$ corresponding to the Sharma-Mittal entropy~(\ref{SM entropy}), the Kaniadakis~entropy (\ref{K-entropy}), 
and the Loop Quantum Gravity entropy (\ref{LQG entropy}) by proceeding to take some limits of the parameters. 

We might consider similar correspondence by using $F(R)$ and Einstein-$F(\mathcal{G})$ gravities. 
In these models, however, the equations corresponding to the first FLRW equation include $\dot H$, 
not as in the entropic equation (\ref{gSeq}) corresponding to the first FLRW equation, which does not include $\dot H$ but only $H$. 
In fact, the equation in $F(R)$ gravity corresponding to the first FLRW equation has the following form, 
\begin{align}
\label{JGRG15}
\frac{F(R)}{6} - \left(H^2 + \dot H\right) F'(R) + 6 \left( 4H^2 \dot H + H \ddot H\right) F''(R)
= \left( \frac{8\pi G}{3} \right) \rho\, .
\end{align}
We should also note that $R=12H^2 + 6\dot H$. 
On the other hand, in the case of the Einstein-$F(\mathcal{G})$ gravity, the corresponding equation is given by 
\begin{align}
\label{GB7b}
H^2 + \left( \frac{8\pi G}{3} \right) \left( F(\mathcal{G}) + \mathcal{G}F'(\mathcal{G}) + 24 \dot{\mathcal{G}}F''(\mathcal{G}) H^3 \right)
= \left( \frac{8\pi G}{3} \right) \rho\, .
\end{align}
In the spatially flat FLRW spacetime~(\ref{metric}), we find $\mathcal{G}= 24 \left( H^2 \dot H + H^4 \right)$. 
Therefore in the cases of $F(R)$ and Einstein-$F(\mathcal{G})$ gravities, the correspondence may exist only in some limit like $|\dot H | \ll H^2$. 
In the limit, Eqs.~(\ref{JGRG15}) and (\ref{GB7b}) have the following forms, 
\begin{align}
\label{JGRG15lim}
\frac{F\left( 12 H^2 \right)}{6} - H^2 F' \left( 12 H^2 \right) =&\, \left( \frac{8\pi G}{3} \right) \rho\, . \\
\label{GB7blim}
H^2 + 64\pi G \left( F \left( 24 H^4 \right) + 24 H^4 F' \left( 24 H^4 \right) \right)
=&\, \left( \frac{8\pi G}{3} \right) \rho\, .
\end{align}
In this limit, one may consider the correspondence between the entropic gravities and $F(R)$ 
and Einstein-$F(\mathcal{G})$ gravities (for review, see~\cite{Nojiri:2010wj, Capozziello:2011et, Nojiri:2017ncd}). 
This could mean, however, that, in the case that the change with respect to time is not small, as in the case of gravitational waves, we cannot use such correspondence. 
Hence, we established the correspondence of $F(T)$ gravity with entropic gravity based on generalised entropy.

\section{Correspondence between $F(Q)$ gravity and generalised entropies}\label{FQrorr}

Besides the $F(T)$ gravity, as an alternative to the gravity theories where metric description is used, the gravity theories based on non-metricity 
have been actively investigated~\cite{Nester:1998mp, BeltranJimenez:2018vdo, Runkla:2018xrv, BeltranJimenez:2019tme, Capozziello:2022tvv}. 
The fundamental geometrical quantity in such theories is a non-metricity scalar $Q$. 
The connection appears as a variable independent from the metric. 
In the theories, we impose conditions that the Riemann tensor and torsion tensor vanish. 
Then the connection can be written by using four scalar fields~\cite{Blixt:2023kyr, BeltranJimenez:2022azb, Adak:2018vzk, Tomonari:2023wcs}. 
These scalar fields are often chosen so that the connection vanishes, which is called the coincident gauge. 
If the action is linear in $Q$, this theory is equivalent to Einstein's gravity because the difference between $Q$ 
and the scalar curvature $R$ is a total derivative. 

As an extension, we may consider the theory where the action is given by a function of $Q$, $F(Q)$ as an analogy 
of $F(R)$ gravity or $F(T)$ gravity. 
In the $F(Q)$ gravity, { there have been long arguments about} the 
dynamical degrees of freedom (DOF)~\cite{Hu:2022anq, DAmbrosio:2023asf, Heisenberg:2023lru, Paliathanasis:2023pqp, Dimakis:2021gby, Hu:2023gui, 
Heisenberg:2023wgk, Gomes:2023tur, Bello-Morales:2024vqk}. 
{ In \cite{Heisenberg:2023wgk}, it has been shown that there are up to seven degrees of freedom. 
After that, in \cite{Gomes:2023tur}, it has been shown that there exist seven degrees of freedom. 
Furthermore, in \cite{Gomes:2023tur}, the existence of the ghost was claimed because the conformal mode of the metric is a ghost. 
Before the paper, however, in \cite{Hu:2023gui}, it had been shown that the conformal mode was a ghost but it does not propagate due to the constraint. 
Therefore the seven degrees of freedom could not be ghosts. 
Recently, in
}
\cite{Nojiri:2024zab}, the authors have proposed to define $F(Q)$ gravity theory by using the metric and only four scalar fields giving the connection as an independent field. 

We now consider $F(Q)$ gravity, whose formulation is given as the following. 
In previous papers, the treatments of the $F(Q)$ gravity are not so well-defined, which is related to the degrees of freedom. 
For example, it is often used the variation of the action with respect to the connection to obtain the corresponding equations. 
In the case of $F(Q)$ gravity, however, the connection is very restricted by requiring that the torsion and the Riemann tensor should vanish. 
Therefore not all the components of the connection are independent. 
One way to solve this problem is to impose the constraints on the connection, that is, the vanishing torsion and the vanishing Riemann curvature, 
by the Lagrange multiplier fields. 
Unfortunately, we have not found any paper where the multiplier fields are solved but often the reduced equations are used. 
In this paper, the $F(Q)$ model is defined only by using the metric and four scalar fields $\xi^a$. 
Therefore, the model under consideration is well-defined. 

The general affine connection on a manifold that is both parallelisable and differentiable can be expressed as follows, 
\begin{align}
\label{affine}
{\Gamma^\sigma}_{\mu \nu}= {{\tilde \Gamma}^\sigma}_{\mu \nu} + K^\sigma_{\;\mu \nu} + L^\sigma_{\;\mu \nu}\,.
\end{align}
Here $\tilde \Gamma^\sigma_{\;\mu \nu}$ is the Levi-Civita connection given by the metric, 
${K^\sigma}_{\mu \nu}$ is the contortion in (\ref{eq:2.4}), and ${L^\sigma}_{\mu \nu}$ is called deformation and is expressed as, 
\begin{align}
\label{deformation}
{L^\sigma}_{\mu \nu}= \frac{1}{2} \left( Q^\sigma_{\;\mu \nu} - Q^{\ \sigma}_{\mu\ \nu} - Q^{\ \sigma}_{\nu\ \mu} \right)\,.
\end{align}
Here ${Q^\sigma}_{\mu \nu}$ is non-metricity tensor defined by, 
\begin{align}
\label{non-metricity}
Q_{\sigma \mu \nu}= \nabla_\sigma g_{\mu \nu}= \partial_\sigma g_{\mu \nu} - {\Gamma^\rho}_{\sigma \mu } g_{\nu \rho} - {\Gamma^\rho}_{\sigma \nu } g_{\mu \rho } \,.
\end{align}
The tensor $Q_{\sigma \mu \nu}$ is used to construct the $F(Q)$ gravity. 

A solution satisfying the condition that both the Riemann tensor and the torsion vanish is given by using four fields $\xi^a$ $\left( a = 0,1,2,3 \right)$, 
\begin{align}
\label{G1B}
{\Gamma^\rho}_{\mu\nu}=\frac{\partial x^\rho}{\partial \xi^a} \partial_\mu \partial_\nu \xi^a \, .
\end{align}
We should note that $\xi^a$'s are scalar fields. 
By using the general coordinate transformation, we often choose the gauge condition ${\Gamma^\rho}_{\mu\nu}=0$, which is called 
the coincident gauge and this condition is realised by the choice of $\xi^a=x^a$. 
The gauge condition, however, often contradicts { metric assumption like that in (\ref{metric}).}

By using the metric and connection, the non-metricity scalar is given by 
\begin{align}
\label{Q}
Q=&\, - \frac{1}{4} g^{\alpha\mu} g^{\beta\nu} g^{\gamma\rho} \nabla_\alpha g_{\beta\gamma} \nabla_\mu g_{\nu\rho}
+ \frac{1}{2} g^{\alpha\mu} g^{\beta\nu} g^{\gamma\rho} \nabla_\alpha g_{\beta\gamma} \nabla_\rho g_{\nu\mu}
+ \frac{1}{4} g^{\alpha\mu} g^{\beta\gamma} g^{\nu\rho} \nabla_\alpha g_{\beta\gamma} \nabla_\mu g_{\nu\rho} \nonumber \\
&\, - \frac{1}{2} g^{\alpha\mu} g^{\beta\gamma} g^{\nu\rho} \nabla_\alpha g_{\beta\gamma} \nabla_\nu g_{\mu\rho} \, .
\end{align}
By using $Q$, the action of $F(Q)$ gravity has the following form, 
\begin{align}
\label{ActionQ}
S= \frac{1}{2\kappa^2} \int d^4 x \sqrt{-g} F(Q)\, .
\end{align}
Because the metric $g_{\mu\nu}$ and $\xi^a$ are regarded as independent fields, we find 
\begin{align}
\label{eq1}
\frac{1}{2\kappa^2} \mathcal{G}_{\mu\nu} \equiv &\, \frac{1}{\sqrt{-g}} g_{\mu\rho} g_{\nu\sigma}\frac{\delta S}{\delta g_{\rho\sigma}} \nonumber \\
=&\, \frac{1}{2} g_{\mu\nu} F(Q)
 - F'(Q) g^{\alpha\beta} g^{\gamma\rho} \left\{ -\frac{1}{4} \nabla_\mu g_{\alpha\gamma} \nabla_\nu g_{\beta\rho}
 - \frac{1}{2} \nabla_\alpha g_{\mu\gamma} \nabla_\beta g_{\nu\rho} \right. \nonumber \\
&\, + \frac{1}{2} \left( \nabla_\mu g_{\alpha\gamma} \nabla_\rho g_{\beta\nu}
+ \nabla_\nu g_{\alpha\gamma} \nabla_\rho g_{\beta\mu} \right)
+ \frac{1}{2} \nabla_\alpha g_{\mu\gamma} \nabla_\rho g_{\nu\beta}
+ \frac{1}{4} \nabla_\mu g_{\alpha\beta} \nabla_\nu g_{\gamma\rho}
+ \frac{1}{2} \nabla_\alpha g_{\mu\nu} \nabla_\beta g_{\gamma\rho} \nonumber \\
&\, \left. - \frac{1}{4} \left( \nabla_\mu g_{\alpha\beta} \nabla_\gamma g_{\nu\rho} + \nabla_\nu g_{\alpha\beta} \nabla_\gamma g_{\mu\rho} \right)
 - \frac{1}{2} \nabla_\alpha g_{\mu\nu} \nabla_\gamma g_{\beta\rho}
 - \frac{1} {4} \left(\nabla_\alpha g_{\gamma\rho} \nabla_\mu g_{\beta\nu} + \nabla_\alpha g_{\gamma\rho} \nabla_\nu g_{\beta\mu} \right)
\right\} \nonumber \\
&\, - \frac{g_{\mu\rho} g_{\nu\sigma}}{\sqrt{-g}} \partial_\alpha \left[ \sqrt{-g} F'(Q) \left\{ - \frac{1}{2} g^{\alpha\beta} g^{\gamma\rho} g^{\sigma\tau} \nabla_\beta g_{\gamma\tau}
+ \frac{1}{2} g^{\alpha\beta} g^{\gamma\rho} g^{\sigma\tau} \left( \nabla_\tau g_{\gamma\beta} + \nabla_\gamma g_{\tau\beta} \right) \right. \right. \nonumber \\
&\, \left. \left. + \frac{1}{2} g^{\alpha\beta} g^{\rho\sigma} g^{\gamma\tau} \nabla_\beta g_{\gamma\tau}
 - \frac{1}{2} g^{\alpha\beta} g^{\rho\sigma} g^{\gamma\tau} \nabla_\gamma g_{\beta\tau}
 - \frac{1}{4} \left( g^{\alpha\sigma} g^{\gamma\tau} g^{\beta\rho} + g^{\alpha\rho} g^{\gamma\tau} g^{\beta\sigma}
\right) \nabla_\beta g_{\gamma\tau}
\right\} \right] \nonumber \\
&\, - \frac{g_{\mu\rho} g_{\nu\sigma}}{\sqrt{-g}} {\Gamma^\rho}_{\alpha\eta} \left[ \sqrt{-g} F'(Q) \left\{ - \frac{1}{2} g^{\alpha\beta} g^{\gamma\eta} g^{\sigma\tau} \nabla_\beta g_{\gamma\tau}
+ \frac{1}{2} g^{\alpha\beta} g^{\gamma\eta} g^{\sigma\tau} \left( \nabla_\tau g_{\gamma\beta} + \nabla_\gamma g_{\tau\beta} \right) \right. \right. \nonumber \\
&\, \left. \left. + \frac{1}{2} g^{\alpha\beta} g^{\eta\sigma} g^{\gamma\tau} \nabla_\beta g_{\gamma\tau}
 - \frac{1}{2} g^{\alpha\beta} g^{\eta\sigma} g^{\gamma\tau} \nabla_\gamma g_{\beta\tau}
 - \frac{1}{4} \left( g^{\alpha\sigma} g^{\gamma\tau} g^{\beta\eta} + g^{\alpha\eta} g^{\gamma\tau} g^{\beta\sigma}
\right) \nabla_\beta g_{\gamma\tau}
\right\} \right] \nonumber \\
&\, - \frac{g_{\mu\rho} g_{\nu\sigma}}{\sqrt{-g}} {\Gamma^\sigma}_{\alpha\eta} \left[ \sqrt{-g} F'(Q) \left\{ - \frac{1}{2} g^{\alpha\beta} g^{\gamma\rho} g^{\eta\tau} \nabla_\beta g_{\gamma\tau}
+ \frac{1}{2} g^{\alpha\beta} g^{\gamma\rho} g^{\eta\tau} \left( \nabla_\tau g_{\gamma\beta} + \nabla_\gamma g_{\tau\beta} \right) \right. \right. \nonumber \\
&\, \left. \left. + \frac{1}{2} g^{\alpha\beta} g^{\rho\eta} g^{\gamma\tau} \nabla_\beta g_{\gamma\tau}
 - \frac{1}{2} g^{\alpha\beta} g^{\rho\eta} g^{\gamma\tau} \nabla_\gamma g_{\beta\tau}
 - \frac{1}{4} \left( g^{\alpha\eta} g^{\gamma\tau} g^{\beta\rho} + g^{\alpha\rho} g^{\gamma\tau} g^{\beta\eta}
\right) \nabla_\beta g_{\gamma\tau}
\right\} \right] \, , 
\end{align}
and 
\begin{align}
\label{eq3}
X_a \equiv&\, \frac{1}{\sqrt{-g}} \frac{\delta S}{\delta \xi^a} \nonumber \\
=&\, - \partial_\sigma
\left[ \frac{\partial x^\eta}{\partial \xi^a} \frac{\partial x^\sigma}{\partial \xi^b}\partial_\xi \partial_\zeta \xi^b \left\{ \sqrt{-g} F'(Q) 
 \left( 
 - \frac{1}{2} g^{\xi\rho} g^{\zeta\nu} g^{\gamma\mu} 
 - \frac{1}{2} g^{\xi\rho} g^{\gamma\nu} g^{\zeta\mu} 
+ 4 g^{\xi\mu} g^{\zeta\nu} g^{\gamma\rho} \right. \right. \right. \nonumber \\
&\, \qquad \qquad \quad \left. \left. \left. + \frac{1}{4} g^{\xi\mu} g^{\zeta\gamma} g^{\nu\rho} 
 - g^{\xi\nu} g^{\zeta\gamma} g^{\mu\rho} 
 - g^{\mu\zeta} g^{\nu\rho} g^{\xi\gamma} 
 - g^{\mu\gamma} g^{\nu\rho} g^{\xi\zeta} 
\right) 
g_{\eta\gamma} \nabla_\mu g_{\nu\rho} \right\} \right] \nonumber \\
&\, - \partial_\xi \partial_\zeta \left[\frac{\partial x^\eta}{\partial \xi^a} \left\{ \sqrt{-g} F'(Q) 
 \left( 
 - \frac{1}{2} g^{\xi\rho} g^{\zeta\nu} g^{\gamma\mu} 
 - \frac{1}{2} g^{\xi\rho} g^{\gamma\nu} g^{\zeta\mu} 
+ 4 g^{\xi\mu} g^{\zeta\nu} g^{\gamma\rho} \right. \right. \right. \nonumber \\
&\, \qquad \qquad \quad \left. \left. \left. + \frac{1}{4} g^{\xi\mu} g^{\zeta\gamma} g^{\nu\rho} 
 - g^{\xi\nu} g^{\zeta\gamma} g^{\mu\rho} 
 - g^{\mu\zeta} g^{\nu\rho} g^{\xi\gamma} 
 - g^{\mu\gamma} g^{\nu\rho} g^{\xi\zeta} 
\right) 
g_{\eta\gamma} \nabla_\mu g_{\nu\rho} \right\} \right] \, .
\end{align}
In principle, Eq.~(\ref{eq3}) is solved with respect to $\xi^a$. 
{ Eq.~(\ref{eq3}) appears in the standard formulation where the Riemann tensor and the torsion tensor vanish by using the 
Lagrange multiplier field \cite{BeltranJimenez:2019tme}. 
Therefore the formulation in \cite{Nojiri:2024zab} could be a simpler version of the standard formulation. 
}

In the spatially flat FLRW spacetime (\ref{metric}), we can solve Eq.~(\ref{eq3}) with respect to $\xi^a$. 
The $\xi^a$ does not, however, appear in the equation (\ref{eq1}) in the spacetime and when we consider the coupling with the matter, 
we obtain an equation corresponding to the first FLRW equation as follows, 
\begin{align}
\label{vQ_FLRW0}
\left( \frac{8\pi G}{3} \right)\rho = \frac{1}{3} \left( \frac{F(Q)}{2} + 6 H^2 F'(Q) \right) \, . 
\end{align}
In the spatially flat FLRW spacetime (\ref{metric}), $Q$ is given by $Q=-6H^2$. 
Therefore the r.h.s. of Eq.~(\ref{vQ_FLRW0}) is given by a function of only $H$ as in Eqs.~(\ref{FTfirst}) and (\ref{gSeq}). 
Because $\frac{F(Q)}{2} + 6 H^2 F'(Q) = \frac{F(Q)}{2} - Q F'(Q) = \left( - Q \right)^\frac{3}{2} \left( \left( - Q \right)^{-\frac{1}{2}} F(Q) \right)'$, 
as in (\ref{gSeqFT}), we find
\begin{align}
\label{gSeqFQ}
\frac{1}{3} \left( \frac{F(Q)}{2} + 6 H^2 F'(Q) \right) = \frac{1}{3} \left( - Q \right)^\frac{3}{2} \left( \left( - Q \right)^{-\frac{1}{2}} F(Q) \right)' 
= \int^{H^2} dx \left. \left(\frac{\partial S_\mathrm{g}}{\partial S}\right) \right|_{S = \frac{\pi}{Gx}} \, ,
\end{align}
and, 
\begin{align}
\label{gSeqFQ2}
F(Q) = 3 \left( - Q \right)^\frac{1}{2} \int^{-\frac{Q}{6}} dy \left( 6 y \right)^{-\frac{3}{2}} 
\int^y dx \left. \left(\frac{\partial S_\mathrm{g}}{\partial S}\right) \right|_{S = \frac{\pi}{Gx}} \, . 
\end{align}
Therefore as in (\ref{fT}), $F(Q)$ gravity theory corresponding to the generalised entropy $S_\mathrm{g}(S)$ is found. 

We should note that the expression (\ref{gSeqFQ2}) is identical to the expression (\ref{fT}) for the $F(T)$ gravity by replacing $T$ with $Q$. 
Therefore we obtain the expressions of $F(Q)$ for the four-parameter generalised entropies in (\ref{intro-1}), 
the Tsallis entropy~(\ref{Tsallis entropy}) or with the Barrow entropy~(\ref{Barrow-entropy}), and the R\'{e}nyi entropy (\ref{Renyi entropy}) 
by replacing $T$ with $Q$ in Eqs.~(\ref{FRW-2FT}), (\ref{HTBFT}), and (\ref{HTRFT}), respectively. 
In some other limits, we can also obtain the expressions of $F(Q)$ corresponding to the Sharma-Mittal entropy~(\ref{SM entropy}), the Kaniadakis~entropy (\ref{K-entropy}), 
and the Loop Quantum Gravity entropy (\ref{LQG entropy}). 

\section{Gravitaional wave}\label{SecGW}

As we mentioned, $F(T)$ gravity is an unphysical model due to the existence of the superluminal propagating modes \cite{Izumi:2012qj, Ong:2013qja}. 
As long as we use the $F(T)$ gravity model as an effective local theory corresponding to the entropic gravity given by the generalised entropies, 
it might not generate any problem perturbatively.
Now we can solve the problem of GW description in entropic gravity. 
For example, the gravitational wave in $F(T)$ gravity may be identified as the gravitational wave in entropic gravity due to the above-established correspondence between the two theories. 
It is well-known that as long as we treat the model perturbatively, only propagating mode in the $F(T)$ gravity
is the gravitational wave corresponding to massless and spin-two mode~\cite{Bamba:2013ooa}. 

On the other hand, in the case of $F(Q)$ gravity theory, any physically crucial problem has not been found different from the $F(T)$ gravity. 
As in the case of $F(T)$ gravity, at least perturbatively, the gravitational wave of massless and spin-two mode is only propagating mode~\cite{Capozziello:2024vix}, 
what is consistent with the observation in \cite{Hu:2023gui} that the scalar mode in the $f(Q)$ gravity does not propagate due to the constraint. 

Let us now consider the perturbation from the flat background $g_{\mu\nu}=\eta_{\mu\nu}$ in $F(T)$ grvity. 
By the perturbation of the vierbein $e^A_\mu$
\begin{align}
\label{vbp1}
e^A_\mu = \delta^A_\mu + \epsilon^A_\mu\, ,
\end{align}
the perturbation of the metric is given by 
\begin{align}
\label{mp1}
g_{\mu\nu} = \eta_{\mu\nu} + h_{\mu\nu}\, , \quad 
h_{\mu\nu} \equiv \delta^A_\mu \eta_{AB} \epsilon^B_\nu + \epsilon^A_\mu \eta_{AB} \delta^B_\nu \, .
\end{align}
Then Eq.~(\ref{eq:2.7}) gives, 
\begin{align} 
\frac{1}{2}F'(0) \left(\partial_\sigma\partial_\nu {{\bar h}^\rho}_{\ \mu} + \partial^\rho \partial_\mu \bar h_{\mu\nu} - \Box \bar h_{\mu\nu} -\eta_{\mu\nu}
\partial_\rho\partial^\sigma {{\bar h}^\rho}_{\ \sigma}\right) 
= \kappa^2 T_{\mathrm{matter}\, \mu\nu}\, .
\label{FEhs}
\end{align}
Here $\displaystyle{\bar h_{\mu\nu} = h_{\mu\nu} - \frac{1}{2}\eta_{\mu\nu} h}$. 
Eq.~(\ref{FEhs}) is the standard equation of the gravitational wave except that the gravitational coupling $\kappa$ is modified by the effective one $\kappa_\mathrm{eff}$ 
as ${\kappa_\mathrm{eff}}^2 = \frac{\kappa^2}{F'(0)}$. 
That is, if $F'(0)=1$, Eq.~(\ref{FEhs}) is identical to the equation of the gravitational wave in Einsten's general relativity. 
This also shows that perturbatively, only the propagating mode is a massless and spin-two gravitational wave 
and any other mode like scalar mode does not appear although the superluminal mode could appear non-perturbatively. 

Even in the case of $F(Q)$ gravity~\cite{Capozziello:2024vix}, by the perturbation in (\ref{mp1}), 
$g_{\mu\nu} = \eta_{\mu\nu} + h_{\mu\nu}$, we obtain the equation identical to (\ref{FEhs}). 
Therefore even in the $F(Q)$ gravity, we obtain the standard equation of the gravitational wave with 
the effective gravitational coupling ${\kappa_\mathrm{eff}}^2 = \frac{\kappa^2}{F'(0)}$. 

{ 
We should note that the difference between $Q$ and the scalar curvature in Einstein's gravity is a total derivative, 
which gives a boundary term. 
The importance of the boundary term was discussed in \cite{Capozziello:2024jir, Capozziello:2024zij} especially for the gravitational wave. 
The boundary term induced the propagating scalar mode as in $F(R)$ gravity. 
}

It is interesting that in the case of the Tsallis entropy~(\ref{Tsallis entropy}) or the Barrow entropy~(\ref{Barrow-entropy}), 
as the expression in (\ref{HTBFT}) tells, $F'(0)$ diverges when $\beta>0$ or $C\neq 0$ or vanishes when $\beta<0$ and $C=0$. 
We should note that $\beta=\delta - 1$ for the Tsallis entropy~(\ref{Tsallis entropy}) and $\beta=\frac{\Delta}{2}$ for the Barrow entropy~(\ref{Barrow-entropy}). 
The large $F'(0)$ means that the effective gravitational coupling ${\kappa_\mathrm{eff}}^2 = \frac{\kappa^2}{F'(0)}$ is small and 
the small $F'(0)$ means that the effective gravitational coupling is large. 
In the early universe, there might be an era when the thermodynamical entropy is described by the Tsallis entropy~(\ref{Tsallis entropy}) or the Barrow entropy~(\ref{Barrow-entropy}). 
In general, the fluctuation of the matter generates the gravitational wave. 
If the effective gravitational coupling ${\kappa_\mathrm{eff}}^2$ is small, the generated gravitational wave could be weak 
but if the effective gravitational coupling ${\kappa_\mathrm{eff}}^2$ is large, the generated gravitational wave could be strong. 
Therefore by observing the primordial gravitational wave, we might be able to check the manifestations of the entropic gravity in the corresponding era. 

\section{Summary and discussion}

In this paper, we considered the correspondence between gravity theories and the several kinds of entropic functions proposed so far. 
The entropic functions define modified cosmology in an analogy with Jacobson~\cite{Jacobson:1995ab} derivation of the Einstein equation from 
the Bekenstein-Hawking entropy~\cite{Bekenstein:1973ur, Hawking:1975vcx} in the framework of thermodynamics. 
We compared the equations, which correspond to the first FLRW equation and are derived by the first law of thermodynamics by using the expressions 
of the several entropies and the Hawking temperature, to the equations, which correspond to the first FLRW equation, in some modified gravities. 
Then we have found that $F(T)$ gravity based on the torsion scalar $T$ and $F(Q)$ gravity based on the non-metricity scalar $Q$ can provide the equations 
identical to those in the entropic cosmology based on the several entropic functions as explicitly shown in (\ref{fT}) and (\ref{gSeqFQ2}). 
Therefore, at least, we found some correspondences between the $F(T)$ and $F(Q)$ gravities and general entropies.

By regarding the corresponding gravity as an effective theory equivalent to the general entropic one, we can consider the local dynamics because the gravity theories 
are described by the local Lagrangian density. 
As one example of the local dynamics, we considered the gravitational wave and we obtained Eq.~(\ref{FEhs}) for both $F(T)$ gravity and $F(Q)$ gravity. 
The obtained equation~(\ref{FEhs}) is identical to that in Einstein's gravity but the gravitational coupling $\kappa$ depends on the functions $F(T)$ and $F(Q)$ 
and the effective gravitational coupling $\kappa_\mathrm{eff}$ is found to be given by ${\kappa_\mathrm{eff}}^2 = \frac{\kappa^2}{F'(0)}$. 
Therefore when $F'(0)=1$, the equation of the gravitational wave in (\ref{FEhs}) is completely identical to that in Einsten's general relativity. 
These results also show that at least perturbatively, only the propagating mode in $F(T)$ and $F(Q)$ gravities and therefore in corresponding entropic analogue theory 
is a massless and spin-two gravitational wave. 
These results may give some hints on the number of the dynamical degrees of freedom in these gravity theories. 

The modification of the gravitational coupling by the effective one is rather general in modified gravity theories. 
Not only $F(T)$ and $F(Q)$ but also $F(R)$ gravity has such property. 
The effective gravitational coupling ${\kappa_\mathrm{eff}}^2 = \frac{\kappa^2}{F'(0)}$ generates a problem which is explicit for Tsallis entropy~(\ref{Tsallis entropy}) 
or the Barrow entropy~(\ref{Barrow-entropy}). 
By Eq.~(\ref{HTBFT}), we find $F'(0)$ diverges (when $\beta>0$ or $C\neq 0$) or vanishes (when $\beta<0$ and $C=0$), 
which corresponds to the vanishing or the divergence of the effective gravitational coupling ${\kappa_\mathrm{eff}}$. 
An epoch when the Tsallis entropy governs the dynamics might appear in the early universe. 
In the early universe, $H^2= \frac{T}{6} = - \frac{Q}{6}$ does not vanish and therefore $F'$ and the effective gravitational coupling could not vanish nor diverge. 
As clear from (\ref{HTBFT}), however, there is a suppression or an enhancement of the effective gravitational coupling by the power of the ratio of $H^2$ and the Planck scale. 
Therefore the generation of the gravitational wave by the fluctuation of matter is also suppressed or enhanced. 
The effective gravitational coupling has some effects on other dynamical phenomena like the merger of some objects, which may also change the fluctuation of the 
cosmic microwave background radiation (CMBR). 
As a final remark, let us mention that it would be interesting to search for similar correspondence of entropic theory 
with modified gravity when black hole thermodynamics and related dynamical effects (shadows, etc) are investigated. 
That will be done elsewhere. 

\section*{ACKNOWLEDGEMENTS}

{ We are indebted in the discussions with Taishi Katsuragawa, Kyosuke Tomonari, and Jose Beltr\'an Jim\'enez.}
This work is supported by the program Unidad de Excelencia Maria de Maeztu CEX2020-001058-M, Spain ( SDO).

\end{document}